\documentclass[twocolumn, nofootinbib, nobibnotes, amsmath,amssymb,aps, pra, floatfix]{revtex4-1}

\usepackage{bbm}
\usepackage{mathtools}
\usepackage[export]{adjustbox}
\usepackage{wrapfig}
\usepackage{import}
\usepackage{bbold}
\usepackage{microtype}
\usepackage{bm} 
\usepackage{graphicx}
\usepackage{dcolumn}
\usepackage{color}
\usepackage{xcolor}
\usepackage{silence}

\renewcommand{\v}[1]{\ensuremath{\mathbf{#1}}} 
\newcommand{\gv}[1]{\ensuremath{\mbox{\boldmath$ #1 $}}} 
\newcommand{\uv}[1]{\ensuremath{\mathbf{\hat{#1}}}} 
\newcommand{\abs}[1]{\left| #1 \right|} 
\newcommand{\pd}[2]{\frac{\partial #1}{\partial #2}} 
\newcommand{\grad}[1]{\gv{\nabla} #1} 
\renewcommand{\div}[1]{\gv{\nabla} \cdot #1} 
\newcommand{\curl}[1]{\gv{\nabla} \times #1} 

\begin{document}

\preprint{APS/123-QED}

\title{Hydrodynamics of nonlinear gauge-coupled quantum fluids}

\author{Y. Buggy}
\author{L.G. Phillips}
 \author{P. {\"O}hberg}
\affiliation{SUPA, Institute of Photonics and Quantum Sciences, Heriot-Watt University, Edinburgh EH14 4AS, United Kingdom}


\begin{abstract}
  By constructing a hydrodynamic canonical formalism, we show that the occurrence of an arbitrary density-dependent gauge potential in the meanfield Hamiltonian of a Bose-condensed fluid invariably leads to nonlinear flow-dependent terms in the wave equation for the phase, where such terms arise due to the explicit dependence of the mechanical flow on the fluid density.
  In addition, we derive a canonical momentum transport equation for this class of nonlinear fluid and obtain an expression for the stress tensor.
  Further, we study the hydrodynamic equations in a particular nonlinear fluid, where the effective gauge potential results from the introduction of weak contact interactions in an ultracold dilute Bose gas of optically-addressed two-level atoms.
  In the Cauchy equation of mechanical momentum transport of the superfluid, two non-trivial terms emerge due to the density-dependent vector potential. 
  A body-force of dilation appears as a product of the gauge potential and the dilation rate of the fluid, while the stress tensor features a canonical flow pressure term given by the inner-product of the gauge potential and the canonical current density.
  By numerical simulation, we illustrate an interesting effect of the nonlinear gauge potential on the groundstate wavefunction of a superfluid in the presence of a foreign impurity.
  We find that the groundstate adopts a non-trivial local phase, which is antisymmetric under reversal of the gauge potential.
  The phase profile leads to a canonical-flow or phase-flow dipole about the impurity, resulting in a skirting mechanical flow.
  As a result, the pressure becomes asymmetric about the object and the condensate undergoes a deformation.
\end{abstract}

\maketitle

\section{\label{sec:intro}Introduction}
In classical mechanics, the interaction of charged particles with the electromagnetic field can be completely described in terms of the force fields $\v{E}$ and $\v{B}$.
The electromagnetic potentials $\phi$ and $\v{A}$ on the other hand, enter merely as auxiliary mathematical quantities bearing no physical significance.
The situation is drastically different in quantum physics: quantisation of a classical theory proceeds from knowledge of the canonical momenta, and it is the energies and momenta which are the central quantities determining the phases of quantum wavefunctions.
As a result, charged particles couple directly to the electromagnetic potentials in the quantum theory, where the form of this coupling notably leads to the Aharanov-Bohm effect and the local gauge invariance of quantum mechanics.
The implications of the fundamental role played by the potentials \cite{aharonov1959significance}, have since led to a diverse range of intriguing physical effects.
These arise through the interplay between particle-particle interactions and applied fields. 
Although the weak field behaviour of gauge-coupled systems is well described by linear response theory, large perturbing field values do not generally allow for a meaningful first order expansion \cite{matt2012}.
As the field is gradually increased, the ordering of the system changes abruptly at certain critical values and a variety of physical phenomena become associated with each intensity range \cite{faze99}: from paramagnetic effects \cite{ston1935}, to the quantum Hall \cite{klitzing1980new,yennie1987integral,pran87} and spin quantum Hall \cite{kane2005quantum,kane2005z,fu2007topological,bernevig2006quantum} effects observed in two-dimensional electron systems.
This notably led to the classification of symmetry protected topological phases of matter \cite{thouless1982quantized,chen2013symmetry} and paved the way for the implementation of topological insulators \cite{qi2011topological,hasan2010colloquium}, illustrating the range of intriguing phenomena which emerge in gauge-coupled many-body systems.\\\\
The charge neutrality of Bose-condensed atomic systems seemingly restricts the discovery of exotic states of matter of this kind.
However, the versatility, controllability and robust character of ultracold quantum gases, have since allowed for the possibility of simulating artificial gauge potentials for charge-neutral systems.
These are generally engineered through combined interactions, such that a system exhibits spatially varying local eigenstates \cite{dalibard2011colloquium,goldman2014light}.
In other words, the action of a gauge potential can be mimicked by imparting a geometric phase onto the wavefunction \cite{berry1984quantal,peskin1989aharonov,dalibard2011colloquium,goldman2014light}.
In this regard, the elucidation of the geometrical nature of the Aharonov-Bohm phase \cite{berry1984quantal} was a landmark in understanding magnetism in quantum mechanics.
Local eigenstates can be induced in a variety of different ways.
Initial attempts exploited the equivalence of the Lorentz and Coriolis forces, by stirring the condensate with a focused laser beam in a magnetic trap \cite{madison2000vortex}, a technique that quickly led to the observation of vortex lattices \cite{abo2001observation}.
More recent implementations have relied almost exclusively on dressing the bare atomic states using light-matter interactions.
For instance, a two-photon Raman scheme \cite{lin2016synthetic} was employed in a series of experiments to engineer both electric \cite{lin2011synthetic} and magnetic \cite{lin2009synthetic} synthetic force fields, as well as synthetic spin-orbit coupling \cite{lin2011spin}, spin Hall effect \cite{beeler2013spin} and partial waves \cite{williams2012synthetic}.
Atomic light-dressing has also opened up the possibility of generating non-Abelian vector potentials with non-commuting components.
These can be implemented for atoms with degenerate eigenstates, and generally emerge when coupling to a laser field produces a degenerate subspace of dressed states \cite{dalibard2011colloquium}.
In addition, efforts have been made to extend the first generation of synthetic potentials - whose space and time dependence are prescribed externally and unaffected by particle motion - and endow these with dynamical properties \cite{lin2016synthetic}. 
For instance, it was shown \cite{edmonds2013} how the introduction of weak collisional interactions in an ultracold dilute Bose gas of optically addressed two-level atoms, gives rise to a nonlinear effective vector potential $\v{A}(\rho)$ acting on the condensate, where $\abs{\v{A}}$ is modulated by the density of the atomic gas. 
Density-dependent gauge potentials have also been proposed \cite{greschner2014density} in spin-dependent optical lattices, by combining periodically modulated interactions and Raman-assisted hopping.
In fact, such a potential was recently experimentally implemented by modulating the interaction strength in synchrony with lattice shaking \cite{PhysRevLett.121.030402}, resulting in a density-dependent hopping amplitude.\\\\
It is the emergence of such nonlinear gauge potentials which has motivated the present study.
From a hydrodynamical point of view this is an interesting situation because the kinetic energy density becomes nonlinear in the fluid density.
Thus, flow depends explicitly on the density profile of the fluid, where the magnitude of flow of a volume element typically increases as the element shrinks.
In this paper, we investigate the fluid stress and body-forces which emerge in a superfluid subject to a nonlinear gauge potential. 
The paper comprises two parts.
In section \ref{sec:canonical_formalism}, we construct a general formalism and demonstrate how the occurrence of an arbitrary density-dependent gauge potential in the meanfield Hamiltonian acting on a macroscopic wavefunction, invariably produces flow-dependent terms in the wave equation for the phase. 
We also derive a canonical momentum transport equation for this class of nonlinear fluid and obtain an expression for the stress tensor.
In section \ref{sec:single_component}, we apply our results from section \ref{sec:canonical_formalism} to a particular type of nonlinear fluid, whose microscopic model is well-established.
We find that the nonlinear gauge potential gives rise to two non-trivial terms in the mechanical momentum transport equation of the superfluid, in the form of a canonical flow pressure and a body-force of dilation.
Finally, by numerical simulation, we illustrate an interesting implication of the nonlinear gauge potential on the ground state phase profile of an inhomogeneous superfluid.
\section{\label{sec:canonical_formalism}General formalism for nonlinear gauge-coupled quantum fluids}
\subsection{Hydrodynamic canonical formalism}
Let us assume that an effective density-dependent vector potential $\v{A}\left( \rho \right)$ and effective scalar potential $\eta\left( \rho \right)$, emerge in the meanfield Lagrangian
\begin{equation}
  L_{MF}=\int d^3\v{r}\left[ \psi^{*}\left( i\hbar\partial_t-\hat{H}_{MF} \right)\psi \right],
  \label{eq:lagrangianMeanfieldSystem}
\end{equation}
governing the dynamics of a macroscopic wavefunction $\psi$.
Since a gauge potential enters the meanfield Hamiltonian, $\hat{H}_{MF}$, in a manner consistent with the minimal substitution, $\hat{\v{p}}\rightarrow\hat{\v{p}}-\v{A}$, our study pertains to the class of quantum fluids described by
\begin{equation}
  \hat{H}_{MF}=\frac{\left( \hat{\v{p}}-\v{A}\left( \rho \right) \right)^2}{2m}+\eta\left( \rho \right)+V\left( \v{r},t \right),
  \label{eq:HamiltonianMeanFieldNonlinearEffectivePotentials}
\end{equation}
where we have included an external potential, $V$.
Further, we shall consider a particular microscopic model leading to a $\hat{H}_{MF}$ of the above form, but presently, let us construct a hydrodynamic formalism for the more general case. 
In accordance with Eq. (\ref{eq:lagrangianMeanfieldSystem}), the Lagrangian density of the nonlinear field, may be presented as 
\begin{multline}
  \mathcal{L}=\frac{i\hbar}{2}\left(\psi^*\dot{\psi}-\dot{\psi}^*\psi\right)\\-\frac{1}{2m}\left[ \left( \hat{\v{p}}-\v{A} \right)\psi \right]^*\left[ \left( \hat{\v{p}}-\v{A} \right)\psi \right]-\rho\eta-\rho V,  
  \label{eq:lagrangianDensityMeanFieldEffectivePotentials}
\end{multline}
where we have denoted partial differentiation with respect to time by a dot and performed the transformation $\mathcal{L}\rightarrow\mathcal{L}-i\hbar\partial_t\left( \psi^*\psi \right)/2$, in order for $\mathcal{L}$ to be real.\\
Perhaps the most conspicuous feature of a density-dependent gauge potential, is the occurrence of a flow nonlinearity in the wave equation for $\psi$.
One may gain insight into the appearance of such a term, by casting the meanfield description into a hydrodynamical form. 
To do so, we write the macroscopic wavefunction in the Madelung \cite{madelung1926}, or polar form
\begin{equation}
  \psi=\sqrt{\rho}e^{i\frac{\theta}{\hbar}},\quad\quad\psi^*=\sqrt{\rho}e^{-i\frac{\theta}{\hbar}},
\label{eq:polar_form_Psi}
\end{equation}
and treat $\rho$ and $\theta$ as the independent field variables.
In terms of these, the Lagrangian density (\ref{eq:lagrangianDensityMeanFieldEffectivePotentials}), assumes the form
\begin{equation}
  \mathcal{L}=-\rho\left( \dot{\theta}+\frac{1}{2}mv^2+\eta+V \right)-\mathcal{Q},
  \label{eq:lagrangianDensityMeanFieldEffectivePotentialsHydrodynamicalForm}
\end{equation}
where, $v=\abs{\v{v}}$, is the magnitude of the mechanical flow, or gauge-covariant flow
\begin{equation}
  \v{v}=\v{u}-\v{A}/m,
  \label{eq:velocity_field_gauge_covariant_polar_form}
\end{equation}
$\v{u}=\grad{\theta}/m$ is the canonical flow, or phase flow, and 
\begin{equation}
  \mathcal{Q}=\frac{\hbar^2}{8m\rho}\left(\grad{\rho}\right)^2,
  \label{eq:quantum_energy_density}
\end{equation}
is a quantum energy density contribution.
We make the distinction between the canonical flow and the gauge flow.
In the canonical flow, we include the total flow which can be accounted for locally by a phase twist in a suitable gauge, whereas the gauge flow denotes the flow contribution from $\v{A}$ which can not be absorbed in the phase without destroying the form of the dynamical equations of the fluid.
The fact that it is not possible to gauge away density-dependent vector potentials, will be covered elsewhere.\\
Since $\mathcal{L}$ is linear in the field velocities and the linear form appears as $-\rho\dot{\theta}$, the field components $\rho$ and $\theta$ play the role of conjugate variables \cite{jackiw1994quantization,buggy2019hydrodynamic}, governed by the canonical field equations
\begin{align}
  \dot{\rho}\left( \v{r} \right)&=\frac{\delta H}{\delta\theta\left( \v{r} \right)}, \label{eq:canonical_equations_hydrodynamic_conjugate_rho}\\
  \dot{\theta}\left( \v{r} \right)&=-\frac{\delta H}{\delta\rho\left( \v{r} \right)}, \label{eq:canonical_equations_hydrodynamic_conjugate_theta}
\end{align}
where the Hamiltonian and Lagrangian densities, are related by
\begin{equation}
  \mathcal{L}=-\rho\dot{\theta}-\mathcal{H},
  \label{eq:LagrangianDensityHamiltonianDensity}
\end{equation}
such that
\begin{equation}
  \mathcal{H}=\rho\left(\frac{1}{2}mv^2+\eta+ V\right) +\mathcal{Q}.
  \label{eq:Hamiltonian_density_nonlinear_quantum_fluid}
\end{equation}
Note that the derivatives on the right hand side of Eqs. (\ref{eq:canonical_equations_hydrodynamic_conjugate_rho}) and (\ref{eq:canonical_equations_hydrodynamic_conjugate_theta}) are functional, or variational derivatives, e.g. $\delta H/\delta\rho=\partial\mathcal{H}/\partial\rho-\grad{}\cdot\left( \partial\mathcal{H}/\partial\left( \grad{\rho} \right) \right)$.
Inserting the Hamiltonian density (\ref{eq:Hamiltonian_density_nonlinear_quantum_fluid}) into the canonical field Eqs. (\ref{eq:canonical_equations_hydrodynamic_conjugate_rho}) and (\ref{eq:canonical_equations_hydrodynamic_conjugate_theta}), yields, respectively, the wave equations
\begin{align}
  \partial_t&\rho+\div{\left( \rho\v{v} \right)}=0, \label{eq:continuity} \\
  \partial_t&\theta+\frac{1}{2}mv^2+\Phi\left( \rho,\v{u} \right)+V+Q=0, \label{eq:QHJgeneralNonlinearEffectivePotentials}
\end{align}
where the density-dependence of the kinetic term within the brackets of Eq. (\ref{eq:Hamiltonian_density_nonlinear_quantum_fluid}) leads to an additional nonlinear flow-dependent term in the wave equation for the phase, such that 
\begin{equation}
  \Phi=-\rho\v{v}\cdot\pd{\v{A}}{\rho}+\eta+\rho\pd{\eta}{\rho},
  \label{eq:nonlinearPotentialQHJE}
\end{equation}
and
\begin{equation}
  Q=-\frac{\hbar^2}{2m}\frac{\nabla^2\sqrt{\rho}}{\sqrt{\rho}},
  \label{eq:quantum_potential}
\end{equation}
is the quantum potential, which emerges due to the quantum energy density $\mathcal{Q}$ in Eq. (\ref{eq:Hamiltonian_density_nonlinear_quantum_fluid}).
Equation (\ref{eq:continuity}) expresses the conservation of mass while Eq. (\ref{eq:QHJgeneralNonlinearEffectivePotentials}) takes the form of a quantum Hamilton-Jacobi equation (QHJE), the gradient of which expresses the conservation of mechanical momentum.\\
In summary, when a fluid is subject to a nonlinear vector potential, the flow $\v{v}$ in Eq. (\ref{eq:velocity_field_gauge_covariant_polar_form}) depends explicitly on the density of the fluid and the kinetic energy density $\kappa=\rho mv^2/2$ becomes nonlinear in $\rho$. 
Thus, the change $\delta\kappa$ in an infinitesimal volume due to $\delta\rho$, is not determined simply by the kinetic energy $mv^2/2$ of the volume as it is typically, since $\delta\kappa=\left(mv^2/2+\rho m\v{v}\cdot\partial\v{v}/\partial\rho\right)\delta\rho$.
As a result, a nonlinear flow-dependent term enters the wave equation for the phase.
This feature is intrinsic to systems whose effective Hamiltonian (\ref{eq:HamiltonianMeanFieldNonlinearEffectivePotentials}) features a density-dependent vector potential $\v{A}\left( \rho \right)$.
\subsection{Canonical momentum transport equation}
In the following section, we derive a canonical momentum transport equation for the nonlinear fluid and investigate the implications of the nonlinear gauge potential on the stress tensor.
To do so, recall that the dynamical state of the matter-field is completely specified by the stress-energy-momentum tensor
\begin{equation}
T_{\mu\nu}=-\sum_{\phi=\rho,\theta}\pd{\mathcal{L}}{(\partial_{\mu}\phi)}\partial_{\nu}\phi +\delta_{\mu\nu}\mathcal{L},
  \label{eq:stress_energy_momentum_tensor}
\end{equation}
while the transport equations governing energy-flow and momentum-flow, follow from the conservation law
\begin{equation}
  \partial_{\mu}T_{\mu\nu}=\partial_{\nu}\mathcal{L},
  \label{eq:conservation_energy_momentum}
\end{equation}
where we have adopted a relativistic-like notation with $\mu=0,1,2,3$. 
The Lagrangian density of the field, is given by Eq. (\ref{eq:lagrangianDensityMeanFieldEffectivePotentialsHydrodynamicalForm}). 
Alternatively, $\mathcal{L}$ may be cast in terms of the fields and their spatial derivatives, by substituting the QHJE (\ref{eq:QHJgeneralNonlinearEffectivePotentials}) into Eq. (\ref{eq:lagrangianDensityMeanFieldEffectivePotentialsHydrodynamicalForm}), which yields 
\begin{equation}
  \mathcal{L}=-\frac{\hbar^2}{4m}\nabla^2\rho+\rho\left( \Phi-\eta \right).
  \label{eq:Lagrangian_multi_component_nonlinear_gauge_potential_polar_form_simplified}
\end{equation}
Rendering $\mathcal{L}$ into this form is essential for evaluating the components of the field stress tensor, $T_{ij}$.
The stress-energy-momentum tensor from Eq. (\ref{eq:stress_energy_momentum_tensor}), characterises the dynamical state of the field by specifying the energy density, the momentum density, and the currents associated with both of these quantities.
The energy density of the field, is
\begin{equation}
  -T_{00}=-\rho\dot{\theta}-\mathcal{L}=\mathcal{H},
  \label{eq:energy_density}
\end{equation}
where the last equality follows from Eq. (\ref{eq:LagrangianDensityHamiltonianDensity}).
The energy current density, takes the form
\begin{equation}
  -T_{k0}=D\dot{\rho}w_k-\rho \dot{\theta}v_k\equiv\mathcal{S}_k, 
  \label{eq:energy_current_density}
\end{equation}
where 
\begin{equation}
  w_k=-\frac{D}{\rho}\nabla_k\rho,
  \label{eq:osmotic_velocity}
\end{equation}
is the osmotic velocity \cite{nelson1966derivation,wyatt2006quantum,faris2014diffusion} and $D=\hbar/\left( 2m \right)$ is the quantum diffusion coefficient. 
The canonical momentum density, reads
\begin{equation}
  T_{0k}=\rho\nabla_k\theta=\rho m u_k\equiv \mathcal{P}_k.
  \label{eq:canonical_momentum_density}
\end{equation}
The canonical momentum current density or stress tensor $T_{jk}$ of the field, is found to be
\begin{equation}
  T_{jk}=\rho m\left( w_jw_k+v_ju_k \right)+\delta_{jk}\mathcal{L}.
  \label{eq:stress_tensor_field}
\end{equation}
Note the distinction between $v_j$ and $u_k$ in the above equation.
The interpretation of Eqs. (\ref{eq:energy_current_density}) and (\ref{eq:stress_tensor_field}) as the respective current densities of the quantities defined in Eqs. (\ref{eq:energy_density}) and (\ref{eq:canonical_momentum_density}), follows from the conservation law in Eq. (\ref{eq:conservation_energy_momentum}), which separates out into an equation of continuity of energy
\begin{equation}
    \partial_t\mathcal{H}+\nabla_i\mathcal{S}_i=-\pd{\mathcal{L}}{t},
  \label{eq:transport_energy}
\end{equation}
and an equation of continuity of momentum
\begin{equation}
  \partial_t \mathcal{P}_k+\nabla_j T_{jk}=\pd{\mathcal{L}}{x_k}.
  \label{eq:transport_momentum}
\end{equation}
Let us cast the above equation into a Cauchy form \cite{aris2012vectors} and describe the momentum transport in the reference frame of the fluid.
Substituting Eqs. (\ref{eq:canonical_momentum_density}) and (\ref{eq:stress_tensor_field}) into (\ref{eq:transport_momentum}) and making use of the continuity of fluid mass, leads to the following \textit{canonical} momentum transport equation in the fluid frame:
\begin{equation}
  m\rho\left( \pd{}{t}+\v{v}\cdot\grad{} \right)u_k=\pd{\mathcal{L}}{x_k}+\nabla_j\Pi_{jk},
  \label{eq:Cauchy_equation_canonical_momentum_1}
\end{equation}
where
\begin{equation}
  \Pi_{jk}=-T_{jk}+m\rho v_ju_k,
  \label{eq:stress_tensor_relation_moving_frame}
\end{equation}
is the fluid stress tensor.
Notice the difference in sign convention used for $\Pi_{jk}$ and $T_{jk}$ in the transport equations (\ref{eq:Cauchy_equation_canonical_momentum_1}) and (\ref{eq:transport_momentum}).
In addition, $\Pi_{jk}$ and $T_{jk}$ differ by a flow-stress term, $m\rho v_ju_k$, as a result of the relative motion between the fluid and field frames.
The $\Pi_{jk}$ define a linear map between the surface normal vectors and the forces acting on these, such that the stress tensor of a fluid may be written in the form \cite{landau1959lifshitz}
\begin{equation}
  \Pi_{jk}=-P\delta_{jk}+\sigma_{jk},
  \label{eq:stress_tensor_pressure_viscocity}
\end{equation}
where $P$ is the fluid pressure associated with normal forces and the $\sigma_{jk}$ account for shearing forces.
Note that $P$ represents the pressure of an infinitesimal volume element which flows with the fluid, and not at a fixed point of space.
As such, for a typical fluid, $P$ is independent of the flow profile of the fluid.
However, for a nonlinear gauge-coupled fluid, this is no longer the case.
Indeed, upon substituting expressions (\ref{eq:Lagrangian_multi_component_nonlinear_gauge_potential_polar_form_simplified}) and (\ref{eq:stress_tensor_field}) into Eq. (\ref{eq:stress_tensor_relation_moving_frame}), we find that the stress tensor of the fluid takes the form of Eq. (\ref{eq:stress_tensor_pressure_viscocity}), where 
\begin{equation}
  \sigma_{jk}=-m\rho w_jw_k,
  \label{eq:quantum_stress_tensor_2}
\end{equation}
is the quantum stress tensor, while the fluid pressure is equivalent to the Lagrangian density of the field from Eq. (\ref{eq:Lagrangian_multi_component_nonlinear_gauge_potential_polar_form_simplified}), such that 
\begin{equation}
  P=-\frac{\hbar^2}{4m}\nabla^2\rho+\rho\left( \Phi-\eta \right).
  \label{eq:pressureNonlinearGaugeCoupledFluid}
\end{equation}
Hence, by virtue of Eq. (\ref{eq:nonlinearPotentialQHJE}), the fluid pressure in a nonlinear gauge-coupled quantum fluid depends explicitly on the flow profile of the fluid.
\section{\label{sec:single_component}The nonlinear gauge-coupled superfluid}
\subsection{The origin of the nonlinear gauge potential}
The formalism outlined in the previous section was general, in the sense that the nonlinear gauge potential was viewed as an arbitrary function of the density.
We now turn our attention to a specific model yielding an effective gauge-coupled Hamiltonian.
In particular, it has been shown \cite{edmonds2013} that the introduction of weak contact-interactions in an optically-addressed dilute Bose gas of two-level atoms, leads to a density-modulated gauge potential acting on the condensate.
To see this, we begin by noting that the microscopic Hamiltonian of the dilute cloud, may be written as 
\begin{equation}
\hat{H}=\hat{H}_{Kin}+\hat{H}_{LM}+\hat{H}_{IP},
\end{equation}
where $\hat{H}_{Kin}$ is the single-particle kinetic energy operator, $\hat{H}_{LM}$ describes the light-matter coupling, which we treat semi-classically, and $\hat{H}_{IP}$ is the interparticle potential
\begin{equation}
  \hat{H}_{IP}=\sum_{i<j}\left(\sum_{a,b=1,2}g_{ab}\delta(\v{r}_i-\v{r}_j)|\lambda_a^{(i)}\lambda_b^{(j)}\rangle\langle \lambda_a^{(i)}\lambda_b^{(j)}|\right)\otimes\mathbb{1}_{ij}.
\end{equation}
Here $i$ and $j$ label the atoms, $|\lambda_a^{(i)}\rangle$ denotes the $a^{th}$ internal state of atom $i$ and $\mathbb{1}_{ij}$ is the identity on the complement of the Hilbert space for particles $i$ and $j$.
The coupling constants are related to the associated scattering lengths, in the customary form $g_{ij}=4\pi\hbar^2 a_{ij}/m$.
We also assume a constant detuning over space, and write
\begin{equation}
    \hat{H}_{LM}=\frac{\hbar\Omega}{2}\sum_j\left(e^{i\phi(\v{r}_j)}|\lambda_1^{(j)}\rangle\langle\lambda_2^{(j)}|+e^{-i\phi(\v{r}_j)}|\lambda_2^{(j)}\rangle\langle\lambda_1^{(j)}|\right)\otimes\mathbb{1}_j,
\end{equation}
where $\Omega$ is the Rabi frequency characterising the light-matter coupling and $\phi$ is the phase of the laser field. 
By assuming that the $N$-body wavefunction is a product of $N$ identical single-particle wavefunctions, we obtain \cite{Butera_2016} a meanfield Lagrangian in the form of Eq. (\ref{eq:lagrangianMeanfieldSystem}), where $\hat{H}_{MF}$ acts on the two-component macroscopic wavefunction, $\psi$, as
\begin{equation}
  \hat{H}_{MF}\psi\left( \v{r} \right)=\left(\frac{\hat{\v{p}}^2}{2m}+\hat{U}_{LM}+\hat{U}_{MF} \right)\psi\left( \v{r} \right)
\end{equation}
where $\hat{U}_{LM}= 
\frac{\hbar\Omega}{2}\left(\begin{array}{cc}
  0 & e^{-i\phi(\v{r})} \\
  e^{i\phi(\v{r})} & 0 \end{array} \right)$, and
$\hat{U}_{MF}=\left(\begin{array}{cc}
g_{11}\rho_1+g_{12}\rho_2 & 0 \\
0 & g_{12}\rho_1+g_{22}\rho_2 \end{array} \right)
$ describes the meanfield collisional effects, with $\rho_i$ representing the density of atoms occupying the $i^{th}$ internal state.
The full wavefunction may be written $\sum_{i=\pm}\psi_i|\chi_i\rangle$, where the $|\chi_{\pm}\rangle$ denote the eigenstates of $\hat{U}_{LM}+\hat{U}_{MF}$, the so-called interacting dressed states.
When the light-matter coupling is much stronger than the interparticle potential, which we will assume to be the case in what follows, these can be approximated by treating $\hat{U}_{MF}$ as a perturbation to $\hat{U}_{LM}$.
Furthermore, by preparing the atoms in a particular dressed state, $|\chi_+\rangle$ say, then, within the adiabatic approximation, we may set $\psi_-$ to zero and obtain a projected meanfield Lagrangian (\ref{eq:lagrangianMeanfieldSystem}), where the meanfield Hamiltonian governing the dynamics of $\psi_+\equiv\psi$, is obtained in the form 
\begin{equation}
  \hat{H}=\frac{\left( \hat{\v{p}}-\v{A} \right)^2}{2m}+W+\frac{\hbar\Omega}{2}+\frac{g}{2}\rho,
  \label{eq:meanfieldHamiltonianProjectedNonlinearGaugePotential}
\end{equation}
where $g=\left( g_{11}+g_{22}+2g_{12} \right)/4$, and the geometric synthetic potentials enter the effective Hamiltonian in the form of a scalar function, $W=\frac{\hbar^2}{2m}|\langle\chi_+|\grad{} \chi_- \rangle|^2$ and a Berry connection $\v{A}=i\hbar\langle\chi_+|\grad{}\chi_+\rangle$.
Notice that the $|\chi_{\pm}\rangle$ depend on $\rho$, and $\v{A}$ inherits this dependence: thus we have a density dependent synthetic gauge potential.
To first order, the synthetic potentials read
\begin{align}
  W&=\frac{\abs{\v{A}^{(0)}}^2}{2m}, \\
  \v{A}&=\v{A}^{\left( 0 \right)}+\v{a}\rho,
  \label{}
\end{align}
where $\v{A}^{0}=-\frac{\hbar}{2}\grad{\phi}$ is the single particle contribution to the vector potential and 
\begin{equation}
  \v{a}=\frac{g_{11}-g_{22}}{8\Omega} \grad{\phi}, 
  \label{eq:effectiveStrengthSingleComponentGaugePotential} 
\end{equation}
controls the effective strength and orientation of the gauge potential.
\subsection{Mechanical momentum transport equation}
The nonlinear gauge-coupled superfluid governed by the effective Hamiltonian from Eq. (\ref{eq:meanfieldHamiltonianProjectedNonlinearGaugePotential}), may be viewed as a particular case of quantum fluid discussed in section \ref{sec:canonical_formalism}, where, in an appropriate gauge, $\eta=g\rho/2$ and $\v{A}=\v{a}\rho$.
Hence, the nonlinear scalar term entering the QHJE (\ref{eq:QHJgeneralNonlinearEffectivePotentials}), now reads
\begin{equation}
  \Phi=-\v{v}\cdot\v{A}+g\rho,
  \label{eq:nonlinearPotentialQHJEDensityModulatedGaugePotential}
\end{equation}
where $\v{v}=\v{u}-\v{a}\rho/m$.
As such, the canonical momentum transport equation for the superfluid is retained in the form of Eq. (\ref{eq:Cauchy_equation_canonical_momentum_1}), where the stress tensor of the fluid takes the form 
\begin{equation}
  \Pi_{jk}=\sigma_{jk}-\delta_{jk}\left( -\frac{\hbar^2}{4m}\nabla^2\rho+\frac{g}{2}\rho^2-\v{J}\cdot\v{A} \right),
  \label{eq:stressTensorFluid}
\end{equation}
where $\v{J}=\rho\v{v}$ and $\sigma_{jk}$ is the quantum stress tensor from Eq. (\ref{eq:quantum_stress_tensor_2}).
Substituting the canonical flow $u_k$ in Eq. (\ref{eq:Cauchy_equation_canonical_momentum_1}) for the mechanical flow $v_k=u_k-a_k\rho/m$, and evaluating $\partial\mathcal{L}/\partial x_k$ holding the fields $\rho$, $\theta$ and their derivatives constant, leads to a Cauchy equation of \textit{mechanical} momentum transport
\begin{equation}
  m\rho \left( \pd{}{t}+\v{v}\cdot\grad{} \right)v_k=\rho f_k + \nabla_j \Pi_{jk},
  \label{eq:Cauchy_equation}
\end{equation}
where $\Pi_{jk}$ is given by (\ref{eq:stressTensorFluid}) and the body-force acting on the superfluid, takes the form
\begin{equation}
  f_k=-\nabla_kV+A_k\div{\v{v}}.
  \label{eq:body_forces_multi_component_gauge_potential}
\end{equation}
Note that in order to obtain the above expression, we have used the continuity equation (\ref{eq:continuity}) and relation $\curl{\v{a}}=\v{0}$.
We have also assumed that $\v{a}$ is independent of time.
Typically, body-forces are associated with external potentials whereas fluid stress is connected with nonlinear potentials.   
However, in the case of a nonlinear vector potential, we see that $\v{A}\left( \rho \right)$ plays a double role in Eq. (\ref{eq:Cauchy_equation}), carrying implications for both $\Pi_{jk}$ and $f_k$.
\subsection{Canonical flow pressure and body-force of dilation}
Since the fluid pressure can be read from the diagonal components of the stress tensor (see Eq. \ref{eq:stress_tensor_pressure_viscocity}), we have
\begin{equation}
  P=-\frac{\hbar^2}{4m}\nabla^2\rho+\frac{g}{2}\rho^2-\v{J}\cdot\v{A}.
  \label{eq:pressure_single_component_nonlinear_fluid}
\end{equation}
Hence the fluid pressure depends on the overlap of the current density and the vector potential, and as such, depends explicitly on the canonical flow, $\v{u}$, of the fluid.
In other words, the fluid pressure becomes a function of both independent dynamical variables $\rho$ and $\v{u}$.
One consequence of this, is that $P$ transforms from one Galilean frame of reference to another.
In order to obtain Galilean covariant transformation laws where the pressure remains an invariant quantity, clearly, the nonlinear potentials will have to be transformed in some fashion.
This will be covered elsewhere.
Expanding the current density in expression (\ref{eq:pressure_single_component_nonlinear_fluid}), the fluid pressure may be written as
\begin{equation}
	P=-\frac{\hbar^2}{4m}\nabla^2\rho+\left( \frac{g}{2}+\frac{a^2}{m}\rho\right) \rho^2-\v{J}_{\v{u}}\cdot\v{A},
  \label{eq:pressure_single_component_nonlinear_fluid_Expanded}
\end{equation}
where $a=\abs{\v{a}}$ and $\v{J}_{\v{u}}=\rho\v{u}$ is the canonical current density.
We shall call the pressure term which depends explicitly on the canonical flow, the \textit{canonical flow pressure}:
\begin{equation}
  P_{\v{u}}=-\v{J}_{\v{u}}\cdot\v{A}.
  \label{eq:canonicalFlowPressureSingleComponent}
\end{equation}
Complementing this pressure term, a nonlinear body-force enters Eq. (\ref{eq:body_forces_multi_component_gauge_potential}) as a result of the time-dependence of $\v{A}\left( \rho \right)$, namely 
\begin{equation}
  f_k^d=A_k\div{\v{v}},
  \label{eq:boddyForceDilation}
\end{equation}
which may be interpreted as a \textit{body-force of dilation}.
This follows from the continuity of fluid mass from Eq. (\ref{eq:continuity}), which can be given the form
\begin{equation}
  \div{\v{v}}=-\frac{1}{\rho}\left( \pd{}{t}+\v{v}\cdot\grad{} \right)\rho.
  \label{eq:continuity_dilation}
\end{equation}
The right hand side of the above equation represents the dilation rate of the fluid \cite{aris2012vectors}.
Therefore, if we track an infinitesimal volume element of fluid as it flows, an additional body-force is exerted throughout the element whenever the size of the volume element changes.
If for instance, the element is compressed as a result of entering an increasing surrounding local pressure field, flow is imparted onto the whole element through $f_k^d$.
This explains how a shrinking infinitesimal volume element acquires additional gauge-flow, which must be the result of a body-force.
\subsection{Ground state canonical flow-dipole in an inhomogeneous superfluid}
\begin{figure}[h]
  \centering
  \includegraphics[width=\columnwidth]{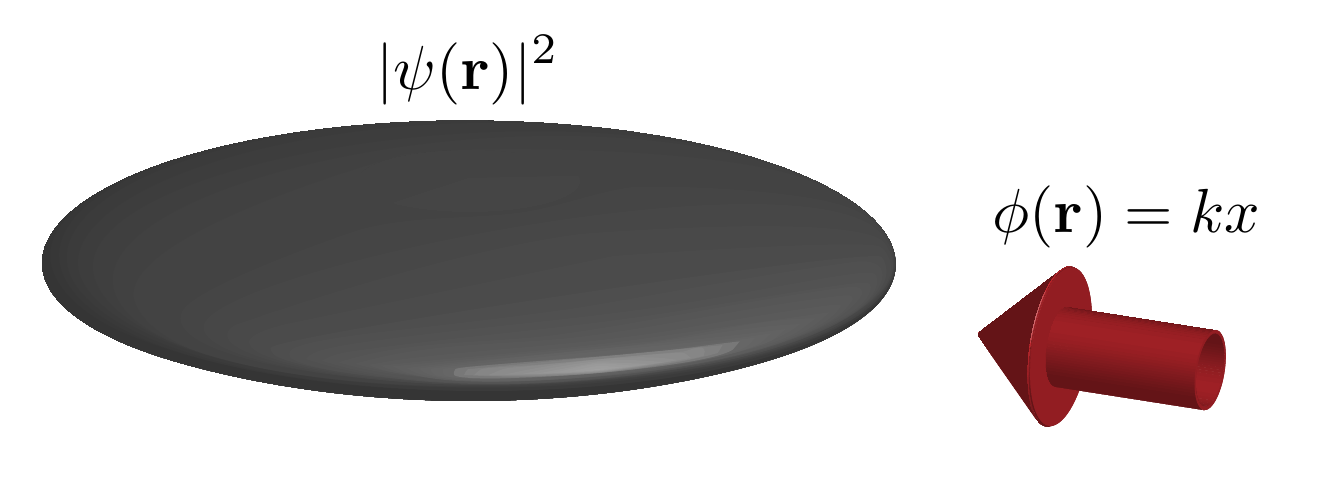}
  \caption{An example of an experimental setup with a superfluid cloud of atoms with density $|\psi({\bf r})|^2$. An incident laser beam with constant intensity and phase $\phi({\bf r})=kx$ where $k$  is the wave number of the light, together with a collisionally induced detuning of the resonant laser frequency, gives rise to the nonlinear gauge potential.}
  \label{experiment}
\end{figure}

In this final section, we illustrate an interesting effect of the nonlinear gauge potential on the ground state wavefunction of an inhomogeneous superfluid.
For the system considered here, the pair of hydrodynamic equations (\ref{eq:continuity}) and (\ref{eq:QHJgeneralNonlinearEffectivePotentials}), are equivalent to the nonlinear Schr\"odinger equation
\begin{equation}
  i\hbar\partial_t\psi=\left[ \frac{\left(\hat{\v{p}}-\v{a}\rho\right)^2}{2m}-\v{a}\cdot\v{J}+g\rho+V \right]\psi,
  \label{eq:GP_equation_single_component_density_modulated_gauge_potential}
\end{equation}
where 
\begin{equation}
  \v{J}=\frac{i\hbar}{2m}\left[ \psi\left( \grad{}+\frac{i}{\hbar}\v{A} \right)\psi^*-\psi^*\left( \grad{}-\frac{i}{\hbar}\v{A} \right)\psi \right],
  \label{eq:currentDensityGaugeCovariant}
\end{equation}
is the gauge-covariant current.
We consider the case of a monochromatic laser field with constant phase twist, {\it i.e.} a plane wave, such that $\v{a}\left( \v{r} \right)$ is constant. See Fig. \ref{experiment} for a description of a possible experimental realisation.
The numerical integration of Eq. (\ref{eq:GP_equation_single_component_density_modulated_gauge_potential}) was achieved using the Crank-Nicholson method for a system of dimension $d=2$ with periodic boundary conditions.
We present results for a condensate populated by $N=1600$ particles in a box with side length $L=47$, comprising $416\times 416$ points.
We let the origin of the system coincide with the center of the box and adopt Cartesian coordinates, denoting the horizontal and vertical axes by $x$ and $y$, respectively.
For our simulation, we have chosen parameters $\abs{\v{a}}=0.73\hbar L^2$ and $g=3.66 \hbar^2 L/\left( 2m \right)$, and set the orientation of the gauge potential at angle $\pi/4$ relative to the $x$-axis, e.g. $\uv{a}=\left( \uv{x}+\uv{y} \right)/\sqrt{2}$.
To establish an inhomogeneous ground state profile, we introduce an immobile impurity into the system, which we model as a Gaussian potential $V\left( \v{r} \right)=20e^{-\abs{\v{r}}^2/2}$.
The ground state was obtained using the method of imaginary time propagation.

\begin{figure}[h]
  \centering
  \includegraphics[width=\columnwidth]{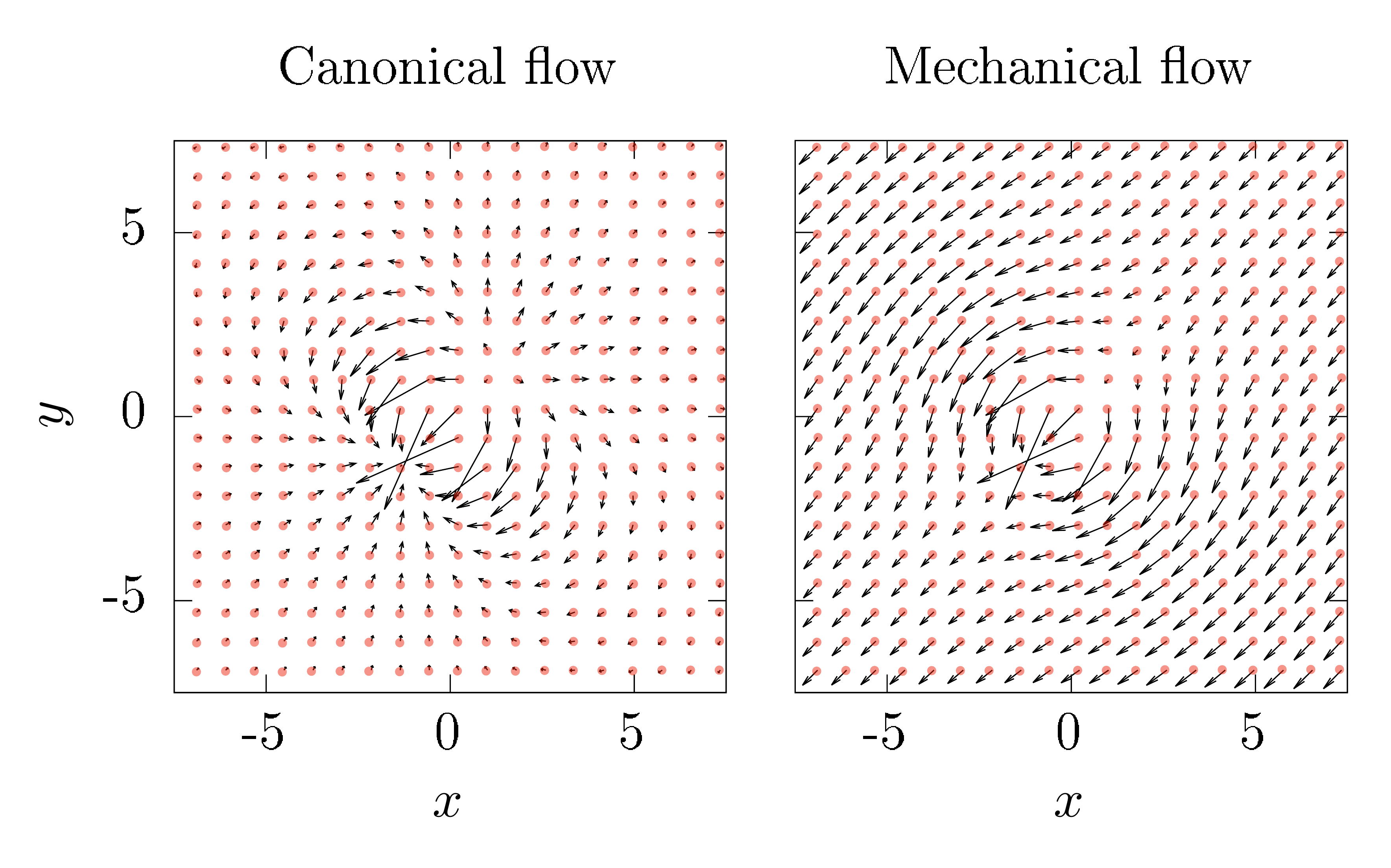}
  \caption{Plots showing the ground state canonical flow, $\v{u}$, and mechanical flow, $\v{v}=\v{u}-\v{A}/m$, in the vicinity of the impurity. The magnitude of the flow in units $\hbar/\left( 2mL \right)$, is given by the arrow length$\times0.068$.}
  \label{fig:GroundStateFlowFields}
\end{figure}

\begin{figure}[h]
  \centering
  \includegraphics[width=\columnwidth]{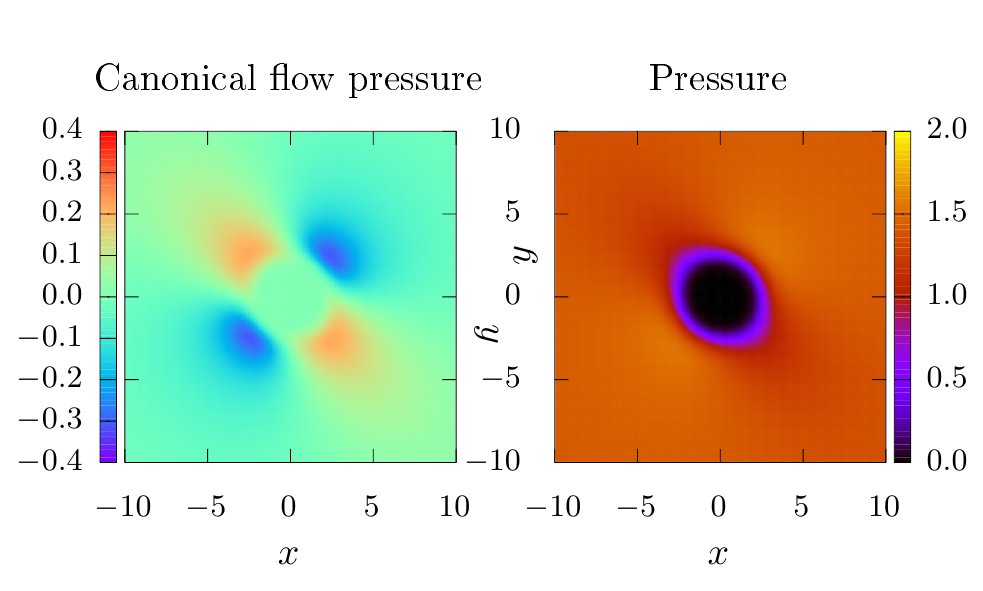}
  \caption{Plots showing the canonical flow pressure, $P_{\v{u}}$, and the fluid pressure, $P$, in the vicinity of the impurity. The pressure is given in units $\hbar^2/\left( 2mL^5 \right)$ and asymmetric about the origin. The impurity is compressed along the axis of the nonlinear gauge ($y=x$) and stretched along the normal axis ($y=-x$).}
  \label{fig:GroundStatePressures}
\end{figure}

\begin{figure}
  \centering
  \includegraphics[width=.8\columnwidth]{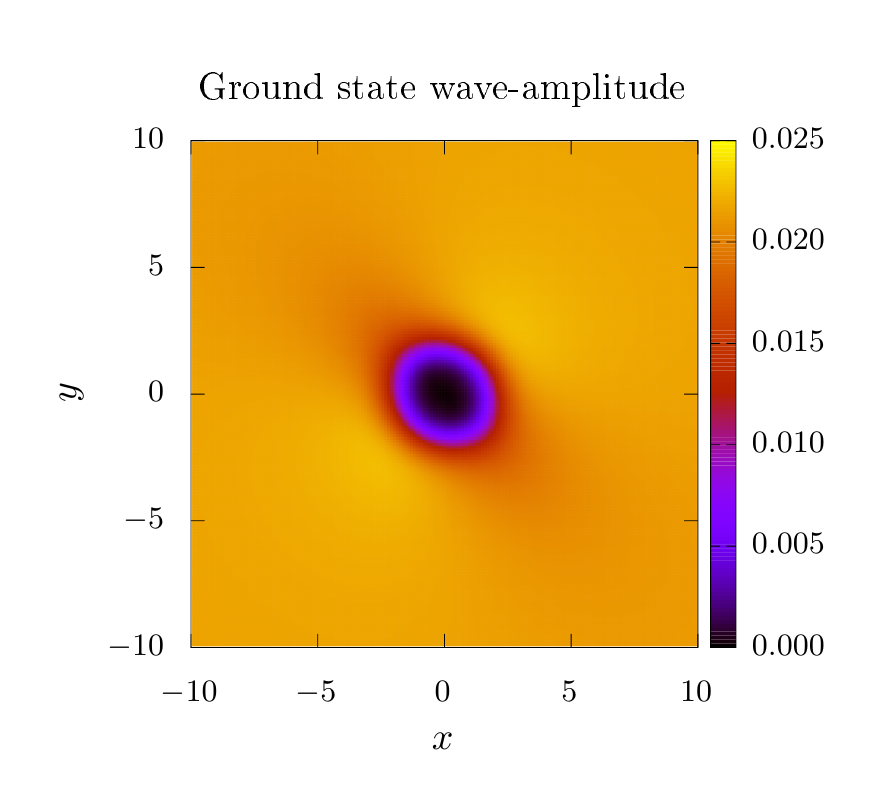}
  \caption{Ground state wave-amplitude, $\abs{\psi}$, in the vicinity of an immobile Gaussian impurity, where $\int d^3\v{r}\abs{\psi}^2=1$.} 
  \label{fig:GroundStateWaveAmplitude}
\end{figure}

In the absence of an impurity, the ground state is that of a homogeneous superfluid, where $\rho=\rho_0$ is constant.
Assuming there is no superflow, the phase is constant over space and oscillates periodically in time. 
However, unlike a standard weakly-interacting superfluid, the gauge-coupled superfluid is not at rest, but exhibits a steady current as a result of the gauge-flow $-\v{a}\rho_0/m$.
In other words, the ground state of the homogeneous superfluid is in a steady state of flow even though no spatial phase twists occur in the system.
Next, let us introduce the localised potential into the system.
This leads to a density-depleted region in the vicinity of the impurity.
As a consequence, the gauge-flow is no longer uniform as in the homogeneous case, but drops in magnitude upon approaching the center of the impurity.
This introduces both non-vanishing transverse and longitudinal components for the gauge-flow $-\v{a}\rho/m$.
The longitudinal component is a significant energy expense for the system, due to the introduction of real-time dependence into the wave-amplitude of the state.
\begin{figure}[h]
  \centering
  \includegraphics[width=\columnwidth]{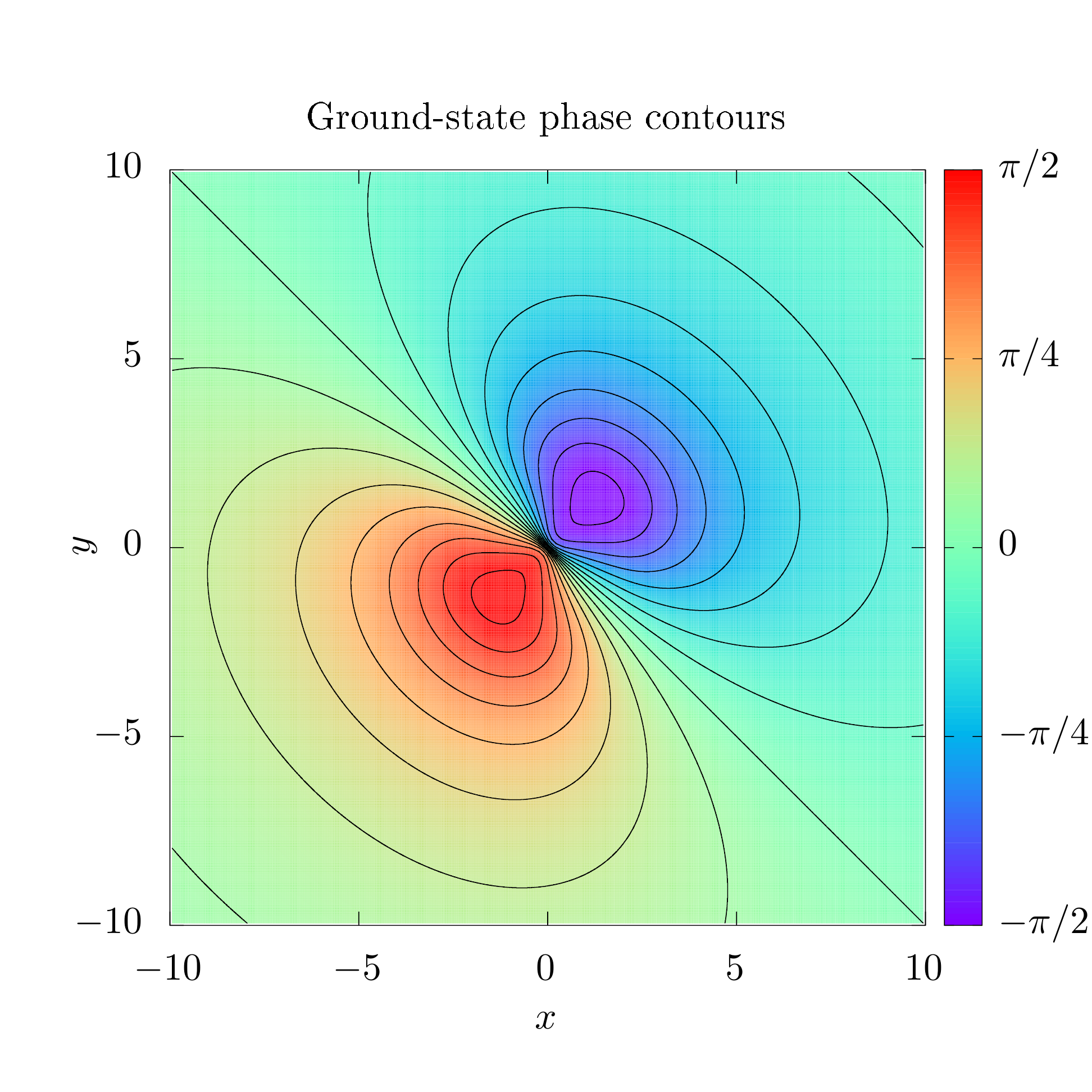}
  \caption{Plot showing the non-trivial phase profile adopted by the ground-state wavefunction in the vicinity of a Gaussian impurity. $17$ contours are included, equally spaced between $-\pi/2$ and $+\pi/2$.}
  \label{fig:GroundStatePhaseContours}
\end{figure}
One may verify numerically that an initial state (with non-vanishing ground state overlap), evolves in imaginary time in such a way that the divergence of the gauge-covariant current approaches zero throughout space.
Notice that in order to achieve this and for the ground state density distribution to be preserved in time, a non-trivial local phase profile must be adopted by the groundstate wavefunction in order to compensate for the non-steady gauge-current.
In other words, the ground state of the system exhibits a non-vanishing canonical flow.
This is illustrated in FIG. \ref{fig:GroundStatePhaseContours}, where we have plotted a series of ground state phase contours in the vicinity of the object, evenly spaced from $-\pi/2$ to $\pi/2$.
The phase is antisymmetric under reversal of the gauge potential, $\v{a}\rightarrow -\v{a}$.
In the bottom left half of the plot, the phase increases from $0$ to $\pi/2$ as we approach $\left( -1.25,-1.25 \right)$, whereas in the upper right half the phase decreases from $0$ to $-\pi/2$ as we approach $\left( 1.25,1.25 \right)$.
In figure \ref{fig:GroundStateFlowFields} we show vector plots of the associated ground-state canonical flow $\v{u}=\left( \psi^*\grad{\psi} -\psi\grad{\psi^*}\right)/\left( i\abs{\psi}^2 \right)$ and mechanical flow $\v{v}=\v{u}-\v{a}\rho/m$.
Here, we notice that $\v{u}$ takes the form of a flow-dipole, leading to a mechanical flow field which skirts around the object.
In turn, the canonical flow-dipole has interesting implications for the fluid pressure.
In figure \ref{fig:GroundStatePressures}, we show the ground state canonical flow pressure and total pressure, computed using expressions (\ref{eq:canonicalFlowPressureSingleComponent}) and (\ref{eq:pressure_single_component_nonlinear_fluid_Expanded}), respectively.
The flow nonlinearity favours occupation (inoccupation) of the blue (red) regions in the left image of FIG. \ref{fig:GroundStatePressures}, leading to an aspherical pressure about the impurity (right image).
This leads to a deformation of the condensate, as seen from the wave-amplitude plot in FIG. \ref{fig:GroundStateWaveAmplitude}.
\section{\label{sec:conclusion}CONCLUSION}
The hydrodynamic canonical formalism is an ideal framework for understanding how a nonlinear gauge potential invariably leads to nonlinear flow terms in the wave equation of a quantum fluid, these resulting from the nonlinear density-dependence of the kinetic energy density.
In turn, two non-trivial terms emerge in the mechanical momentum-transport equation of a superfluid subject to a density-modulated gauge potential, in the form of a canonical flow pressure and a body force of dilation.
The nonlinear gauge potential has interesting implications for an inhomogeneous superfluid, where a nontrivial local phase is adopted by the ground state wavefunction.
The nonlinear gauge potential also has important consequences for the Galilean covariance of the fluid, where new transformations laws are required in order to restore the invariance of the fluid under the transformation group.
The canonical flow pressure should also carry significant implications for the elementary excitations of the fluid.
For instance, one should no longer expect the velocity of sound to be determined exclusively by the adiabatic compressibily, since $P$ depends explicitly on the flow.
This calls for a generalised expression relating the velocity of sound to the fluid pressure.
Finally, the nonlinear body force of dilation should appear in the drag force acting on a moving impurity and may be investigated numerically.
For typical quantum fluids, the drag force is determined by the configuration of the fluid density in the vicinity of the localised object potential.
In contrast, the reaction to the body force of dilation should occur throughout the whole fluid, taking place wherever the divergence of the velocity field is non-vanishing.
Here, we would expect the onset of vortex nucleation to depend on the relative orientation of the gauge potential with respect to the travelling impurity. 

\begin{acknowledgements}
We would like to thank Manuel Valiente for helpful and interesting discussions.
Y.B acknowledges support from EPSRC CM-CDT Grant No. EP/L015110/1.
L.G.P acknowledges support from the EPSRC CM-CDT.
P.{\"O} acknowledges support from EPSRC grant No. EP/M024636/1.
\end{acknowledgements}

\bibliographystyle{apsrev4-1}

%

\end{document}